\documentclass[twocolumn,english,aps,pra,superscriptaddress]{revtex4-1}
\usepackage[T1]{fontenc}
\usepackage[latin9]{inputenc}
\usepackage{amsmath}
\usepackage{amssymb}
\usepackage{graphicx, color, bbm, bbold}

\makeatletter

\newcommand*\LyXThinSpace{\,\hspace{0pt}}

\@ifundefined{textcolor}{}
{%
 \definecolor{BLACK}{gray}{0}
 \definecolor{WHITE}{gray}{1}
 \definecolor{RED}{rgb}{1,0,0}
 \definecolor{GREEN}{rgb}{0,1,0}
 \definecolor{BLUE}{rgb}{0,0,1}
 \definecolor{CYAN}{cmyk}{1,0,0,0}
 \definecolor{MAGENTA}{cmyk}{0,1,0,0}
 \definecolor{YELLOW}{cmyk}{0,0,1,0}
}


\newcommand{\ketbra}[2]{| #1 \rangle \langle #2 |}
\newcommand{\ket}[1]{| #1 \rangle}

\newcommand{\tr}{\mathrm{tr}}

\newcommand{\1}{{\rm 1\hspace{-0.9mm}l}}

\newtheorem{definition}{Definition}

\makeatother

\usepackage{babel}
\begin{document}

\title{Stabilizable Gaussian states}

\date{\today}

\author{\L ukasz Rudnicki}
\email{rudnicki@cft.edu.pl}

\selectlanguage{english}%

\affiliation{Center for Theoretical Physics, Polish Academy of Sciences, Aleja
Lotnik{\'o}w 32/46, 02-668 Warsaw, Poland}

\author{Clemens Gneiting}
\email{clemens.gneiting@riken.jp}

\affiliation{Theoretical Quantum Physics Laboratory, RIKEN, Wako-shi, Saitama
351-0198, Japan }
\begin{abstract}
The unavoidable interaction of quantum systems with their environment usually results in the loss of desired quantum resources. Suitably chosen system Hamiltonians, however, can, to some extent, counteract such detrimental decay, giving rise to the set of stabilizable states. Here, we discuss the possibility to stabilize Gaussian states in continuous-variable systems. We identify necessary and sufficient conditions for such stabilizability and elaborate these on two benchmark examples, a single, damped mode and two locally damped modes.  The obtained stabilizability conditions, which are formulated in terms of the states' covariance matrices, are, more generally, also applicable to non-Gaussian states, where they may similarly help to, e.g., discuss entanglement preservation and/or detection up to the second moments.
\end{abstract}
\maketitle
\section{Introduction}

Quantum control is a model example of research pursued in modern physics,
in which theoretical or even deeply mathematically oriented questions
are very close to experimental needs and practical issues. Every experimental
scenario has to take into account its environment, which we are forced
to treat as an independent entity. Clearly, one can make efforts to
isolate a given setup from the influence of the environment. Still,
there is a level at which current technological possibilities become
exhausted. While in the classical domain one can usually allow for
moderately limited interactions between the system and its environment,
in the quantum case the restrictions become much more severe. Among
particularly relevant examples of that kind are fault-tolerant quantum
computation \cite{FTQC} and the finally successful detection of gravitational
waves \cite{Grav}.

The theory of quantum control offers an alternative approach to the
problem. As one must accept some interactions with the environment,
their undesired consequences may at least be reduced,
if not neutralized. To be more concrete, in many experimental scenarios one
has additional freedom in terms of tunable external interactions admitted by the setup,
given in terms of a so-called control Hamiltonian. Given a control objective,
the question posed in (optimal) quantum control thus refers to the best selection
and tuning of such a Hamiltonian, in order to minimize the detrimental
influence of the environment. Objectives may, for instance, be the preservation
of desired quantum resources, such as coherence or entanglement.

Given a family of quantum states, knowledge about the environment and the functional
form of implementable interactions (these may, e.g., be experimentally
imposed constraints and knowledge), one can ask whether and to which
extent these ingredients are able to interplay, to cooperate. Of particular
interest are families of states which can be fully protected against the
environment, so-called stabilizable states. These stabilizable states then set
the benchmark to what extent the control objectives can be achieved. 
As it turns out, in many relevant cases, structural properties of quantum mechanics allow us to gain
unexpectedly general insights into the composition of the set of stabilizable
states,  potentially rendering ``brute-force'' optimizations unnecessary.

In this article, we focus on Gaussian states. Gaussian states play a central role
in quantum information theory and in particular in quantum optics. On the one hand, this is due
to their natural occurrence and thus experimental availability in quantum optical settings; think
only of coherent states \cite{Mandel1995optical, Walls2007quantum}. On the other hand,
they have been identified as essential resources for many quantum information and communication
protocols in continuous variables \cite{Wang2007quantum, Weedbrook2012gaussian}, including,
e.g., teleportation \cite{Braunstein1998teleportation, Furusawa1998unconditional, Liuzzo2017optimal}.
Moreover, while most of their applications refer to the light field, Gaussian states also represent
important resources in the case of matter waves, e.g., to detect entanglement in matter wave
interference \cite{Kheruntsyan2005einstein, Gneiting2013nonlocal}.

\subsection{Stabilizable states} We are interested in an open-system
dynamics of a quantum system governed by the equation ($\hbar=1$)
\begin{equation}
\dot{\varrho}=i\left[\varrho,\hat{H}\right]+\mathcal{D}\left(\varrho\right),\label{evol}
\end{equation}
with $\hat{H}$ being the (control) Hamiltonian and $\mathcal{D}\left(\varrho\right)$
representing the ``dissipative'' part of the time evolution. Given the
evolution equation (\ref{evol}), we introduce and discern two similar types of quantum states:
\begin{definition}[Stationary states]\label{def1}
A state $\varrho$ is a stationary state of the master equation (\ref{evol}), if $\dot{\varrho}=0$.
\end{definition}
\begin{definition}[Stabilizable states]\label{def2}
A state $\varrho$ is a stabilizable state with respect to the dissipator $\mathcal{D}$, if there exists a Hamiltonian $\hat H$ such that $\varrho$ is the stationary state of the master equation (\ref{evol}) with this specific Hamiltonian as an input.
\end{definition}
The latter definition allows us to introduce the {\it set of stabilizable states} as
\begin{equation}
\mathcal{S}=\left\{ \varrho:\quad\exists_{\hat{H}}\;\textrm{such that }\:i\left[\varrho,\hat{H}\right]+\mathcal{D}\left(\varrho\right)=0\right\} .\label{stabilizability}
\end{equation}
While stationary and stabilizable states are, of course, related notions, the fundamental difference between them lies in that the stationary states refer to both a given Hamiltonian and a given dissipator, while the stabilizable states refer to a given dissipator only, {\it leaving the Hamiltonian open}. The stabilizable states are thus more general in that they allow one to address the principal stabilization possibilities/limitations of a given dissipator w.r.t. {\it any} Hamiltonian. This will become apparent below, as our conditions for stabilizability are formulated without reference to a Hamiltonian. In the current contribution we are concerned with the stabilizable states.

In principle, the Hamiltonian entering Definition \ref{def2} could be arbitrary, as it may be useful, e.g., for general considerations \cite{Sauer}. For practical or formal reasons, it may, however, be desirable to restrict the class of admitted Hamiltonians. Below, for instance, we will focus on Gaussian (quadratic) Hamiltonians only.

For a given Hamiltonian, one can examine stabilizability of a state $\varrho_0$ by definition,
namely by checking if the conditions
\begin{equation}
i\left\langle \psi_{m}\right|\left[\hat{H},\varrho_0\right]\left|\psi_{n}\right\rangle =\left\langle \psi_{m}\right|\mathcal{D}\left(\varrho_0\right)\left|\psi_{n}\right\rangle ,\label{stabDef}
\end{equation}
are satisfied for all pairs of indices $\left(m,n\right)$, with $\left\{ \left|\psi_{n}\right\rangle \right\} $
being a complete set of states in an underlying Hilbert space. However,
as already mentioned, it is possible to do more than testing Eq. \ref{stabDef}.
Since the time evolution induced by the Hamiltonian part is \emph{unitary},
there are quantities, namely eigenvalues of $\varrho$ or equivalently
all its moments $\textrm{Tr}\varrho^{n}$, which are not sensitive
to it. Thus, if $\mathcal{D}\left(\varrho\right)$ modifies the spectrum
(moments), there is no hope to satisfy (\ref{stabDef}), regardless
of the Hamiltonian. In other words, some dissipators, considered together
with particular quantum states, can admit control Hamiltonians rendering
stabilizability, while other configurations do not allow for that.
A priori, one might not expect that the problem of searching for the optimal Hamiltonian can be tackled independently of the Hamiltonian.

More formally, one can find that \cite{Sauer}
\begin{equation}
\frac{d}{dt}\textrm{Tr}\varrho^{n}=n\textrm{Tr}\left[\mathcal{D}\left(\varrho\right)\varrho^{n-1}\right],\label{General}
\end{equation}
because, due to the cyclic property of the trace, we have $\textrm{Tr}\left(\left[\varrho,H\right]\varrho^{n-1}\right)=0$.
In other words,
\begin{equation}
\varrho\in\mathcal{S}\Longrightarrow\forall_{n\geq2}\;\textrm{Tr}\left[\varrho^{n-1}\mathcal{D}\left(\varrho\right)\right]=0,\label{crit1}
\end{equation}
where the case $n=1$ is excluded, as it is trivial due to normalization
of $\varrho$. Moreover, if $\varrho$ has a non-degenerate spectrum,
the above necessary criteria are also sufficient \cite{Sauer}, as
will later be explained in detail.

Let us emphasize that, in contrast to a {\it per pedes} determination of the set of stabilizable states (for a given dissipator) indirectly via the stationary states, which would require to parametrize the most general Hamiltonian for the considered system, the stabilizability conditions (\ref{General}) allow us to characterize/narrow down the stabilizable states in a geometric, systematic, and hierarchical way.

The conditions (\ref{crit1}) have proven to be well suited to discuss stabilizability in the context of finite-dimensional quantum systems. For example, it could be shown that, in the generic case of two qubits suffering from local amplitude damping, the possible entanglement which can be uphold by even arbitrary interacting Hamiltonians is severely limited \cite{Sauer, Sauer2014stabilizing}.

We note that, for conceptual clarity, in the stabilizability-related considerations the environmentally-induced dissipative part of the system dynamics is assumed to be independent of the chosen Hamiltonian; whereas in microscopic derivations the Hamiltonian in general depends on the interaction with the environment. While such assumption is unproblematic in the case of Markovian master equations, non-Markovian cases must be treated with care \cite{Addis2016problem}. In this article, we only discuss Markovian master equations.

\subsection{Stabilizability in continuous variables} 
\label{Sec1B}
The aim of the current contribution is to generalize stabilizability to continuous variables states relevant for quantum optics. The emphasis is put on Gaussian states; however, the approach presented here can be extended to a more general setting, as outlined below.

In the special case of a single mode, described by the annihilation operator
$\hat{a}$ and the dissipator
\begin{equation} \label{Eq:damping_dissipator}
\mathcal{D}_{a}\left(\varrho\right)=\gamma\hat{a}\varrho\hat{a}^{\dagger}-\frac{\gamma}{2}\left\{ \hat{a}^{\dagger}\hat{a},\varrho\right\} ,
\end{equation}
one can easily see that all coherent states $\left|\alpha\right\rangle $
are stabilizable. This property already follows from the sole
definition, as the linear Hamiltonian $\hat{H}_{\alpha}=i\gamma\left(\alpha\hat{a}^{\dagger}-\alpha^{*}\hat{a}\right)/2$
solves the equation in (\ref{stabilizability}). For $\varrho_{\alpha}=\left|\alpha\right\rangle \left\langle \alpha\right|$
we find
\begin{equation}
\frac{\mathcal{D}_{a}\left(\varrho_{\alpha}\right)}{\gamma}=-\frac{i}{\gamma}\left[\varrho_{\alpha},\hat{H}_{\alpha}\right]=\left|\alpha\right|^{2}\varrho_{\alpha}-\frac{\alpha\hat{a}^{\dagger}\varrho_{\alpha}+\alpha^{*}\varrho_{\alpha}\hat{a}}{2}.
\end{equation}
For this special case also the criteria (\ref{crit1}) give an
affirmative answer: Since $\varrho_{\alpha}^{n}=\varrho_{\alpha}$, we
get a single condition which can easily be shown to be satisfied.
However, as pure coherent states are highly degenerate, we still
need to \emph{guess} the right Hamiltonian.

Stabilizability of quantum states is a notion, which, if viable,
can also provide relevant geometrical intuition \cite{Sauer}.
However, when dealing with continuous variables, the problem of deciding
whether a generic quantum state is stabilizable or not becomes in general intractable
due to the infinite dimension of the Hilbert space. In particular, one
would need to verify infinitely many conditions in the hierarchy (\ref{crit1}).
Our aim is thus to reduce the problem to a manageable size, by utilizing
the properties of the states of continuous variables. Focusing on Gaussian states, we develop, based on the covariance matrix, criteria applicable
to multimode fields. In more general terms
we follow the strategy that, whenever the Hamiltonian part of
the dynamics acts in a ``restricted'' way, one can formulate stabilizability
by considering quantities which are invariant with respect to this
action. In the most general case, we deal with unitary transformations; in specific scenarios, however, the geometry of the problem may support more symmetry. For Gaussian states, in particular, the Hamiltonian time evolution
is realized through symplectic matrices, though one can consider the
action of the symplectic group.

We remark that the problem of optimal control for Gaussian
states has already been successfully studied \cite{Mancini,Mancini2},
also in the context of pure-state preparation \cite{Yamamoto1,Yamamoto2}.
In the current perspective, the emphasis lies on the
geometrical side of the problem, accurately grasped by the notion of the
set of stabilizable states.

\section{Time evolution of Gaussian states}

We consider the general case of an $N$-mode system, described by a
$2N$-dimensional vector \cite{Serafini}
\begin{equation}
\hat\xi=\left(\hat x_{1},\ldots,\hat x_{N},\hat p_{1},\ldots,\hat p_{N}\right)^{T}.
\end{equation}
The position and momentum variables $\hat x$ and $\hat p$ correspond to quadratures of the electromagnetic field. In another experimental context they would have other
meanings, for instance, they could describe the transverse degrees of freedom of photons generated in spontaneous parametric down conversion \cite{SPDC}.

The coherent part of the dynamics is assumed to be linear, i.e.,
it is described by quadratic Hamiltonian \cite{Serafini}
\begin{equation}
\hat{H}=\frac{1}{2}\hat{\xi}^{T}G\hat{\xi},\label{HamiltonianG}
\end{equation}
with $G$ being a real, symmetric matrix. The part responsible for
decoherence effects is expressed in Lindblad form 
\begin{equation}
\mathcal{D}\left(\varrho\right)=\sum_{k=1}^{M}\left(\hat{L}_{k}\varrho \hat{L}_{k}^{\dagger}-\frac{1}{2}\left\{\hat{L}_{k}^{\dagger}\hat{L}_{k},\varrho\right\} \right),\label{DissipatorG}
\end{equation}
with the Lindblad operators $\hat{L}_{k}=c_{k}^{T}\hat{\xi}$ also being linear
in the phase-space variables. By $c_{k}$, following \cite{Yamamoto1,Mancini2},
we denote the $2N$-dimensional complex vectors.

Every Gaussian state $\varrho$ is completely characterized by the
vector of mean values $\xi= \tr (\hat{\xi} \varrho)  $ and the covariance
matrix $V = \tr (\hat{V} \varrho)  $, with  $\hat V = \frac{1}{2}(\Delta \hat{\xi} \Delta \hat{\xi}^T + \mathrm{h.c.})$ and $\Delta \hat{\xi} = \hat{\xi} - \xi$. The evolution equation (\ref{evol}), with inputs (\ref{HamiltonianG})
and (\ref{DissipatorG}), translated to these degrees of freedom reads
\cite{Yamamoto1,Mancini2}:\begin{subequations}
\begin{equation}
\frac{d}{dt}\xi=A\xi ,\label{shift}
\end{equation}
\begin{equation}
\frac{d}{dt}V=AV+VA^{T}+J\left(\textrm{Re}C^{\dagger}C\right)J^{T},\label{CovarEq}
\end{equation}
with 
\begin{equation}
A=J\left[G+\textrm{Im}C^{\dagger}C\right],
\end{equation}
and  $C=\left(c_{1},\ldots,c_{M}\right)^{T}$ being a $M\times2N$
complex dissipation matrix. \end{subequations}By 
\begin{equation}
J=\left(\begin{array}{cc}
0 & \1_{N}\\
-\1_{N} & 0
\end{array}\right),
\end{equation}
we denote a defining matrix of the symplectic group $Sp(2N,\mathbb{R})$.
Note that $J^{-1}=J^{T}=-J$, and $J^{2}=-\1_{2N}$. The above description of time evolution remains valid (though not complete) for all non-Gaussian states.

Finally, we briefly examine the properties of the matrix
$C^{\dagger}C$. Since $C_{jk}=\left(c_{j}\right)_{k}$ with $\left(c_{j}\right)_{k}$
being the $k$-th coordinate of $c_{j}$, we obtain: \begin{subequations}
\begin{equation}
\left(\textrm{Re}C^{\dagger}C\right)_{kl}=\sum_{j=1}^{M}\frac{\left(c_{j}\right)_{k}^{*}\left(c_{j}\right)_{l}+\left(c_{j}\right)_{k}\left(c_{j}\right)_{l}^{*}}{2},
\end{equation}
\begin{equation}
\left(\textrm{Im}C^{\dagger}C\right)_{kl}=\sum_{j=1}^{M}\frac{\left(c_{j}\right)_{k}^{*}\left(c_{j}\right)_{l}-\left(c_{j}\right)_{k}\left(c_{j}\right)_{l}^{*}}{2i},
\end{equation}
so that
\begin{equation}
\left(\textrm{Im}C^{\dagger}C\right)^{T}=-\textrm{Im}C^{\dagger}C,\quad\left(\textrm{\textrm{Re}}C^{\dagger}C\right)^{T}=\textrm{Re}C^{\dagger}C.\label{ImRe}
\end{equation}
\end{subequations}

\section{Symplectic eigenvalues and necessary criteria for stabilizability of Gaussian states}

The coherent evolution of Gaussian states, as governed by the
matrix $G$, can always be represented in the $Sp(2N,\mathbb{R})$
group. In other words, the coherent evolution does not affect symplectic
eigenvalues of $V$ being the eigenvalues of the matrix $\tilde{V}=JV$.
When evaluating the time derivative of the moments of $\tilde{V}$,
we indeed find
\begin{equation}
\frac{d}{dt}\textrm{Tr}\tilde{V}^{n}=n\textrm{Tr}\left(Q\tilde{V}^{n-1}\right),\label{tilde}
\end{equation}
with
\begin{equation}
Q=-\left[\tilde{V},GJ\right]+\left\{ \tilde{V},\left(\textrm{Im}C^{\dagger}C\right)J\right\} +\left(\textrm{Re}C^{\dagger}C\right)J.\label{Q}
\end{equation}
To obtain (\ref{Q}) one starts from (\ref{CovarEq}) and
simplifies using the properties of $J$ and Eq. \ref{ImRe}. 

Similarly to the general case (\ref{General}), the commutator part
of $Q$ does not contribute to the trace in (\ref{tilde}), so that
stabilizability of the state in question is independent of the Hamiltonian
represented here by the matrix $G$. Finally, we find that, if the
Gaussian state $\varrho$ is stabilizable, then
\begin{equation}
2\textrm{Tr}\left[\left(\textrm{Im}C^{\dagger}C\right)J\tilde{V}^{n}\right]+\textrm{Tr}\left[\left(\textrm{Re}C^{\dagger}C\right)J\tilde{V}^{n-1}\right]=0,\label{CriteriaMain}
\end{equation}
for all $n=1,\ldots,2N$. Note that, in this case, we start with $n=1$,
as normalization of $\varrho$ is not related to the trace of $\tilde{V}$.

To fulfill the definition (\ref{stabilizability}) rigorously, one
also needs to require that $\xi=0$, as
otherwise, by virtue of (\ref{shift}), the center of the Gaussian
would move in time. On the other hand, the approach presented above
offers us a possibility to relax the notion in question: One may describe the set of Gaussian states with stationary covariance matrix,
$\dot{V}=0$, irrespective of the evolution of the mean value.

\section{Sufficiency of the stabilizability criteria} 

The criteria (\ref{CriteriaMain}) presented above are by construction
necessary. From the perspective of stabilizability, it is
interesting to strengthen these results to obtain unambiguous conclusions.
To this end, in the general case of Eq. \ref{crit1}, it is essential
to have a non-degenerate spectrum of the whole density matrix. In our case, however,
since the dimension of the Hilbert space is infinite, this requirement
is in general not convenient. On the other hand, the criteria (\ref{CriteriaMain}), established
by means of the covariance matrix, nurture the hope that their sufficiency
can be obtained under significantly milder assumptions.

We start by defining the matrix $\mathcal{V}=V^{1/2}JV^{1/2}$. This
matrix is skew-symmetric (hence also normal) and has an eigendecomposition:
$\pm i\gamma_{j}$, $\left|\gamma_{j}^{\pm}\right\rangle $, for $j=1,\ldots,N$.
The $\gamma_{j}$'s are the symplectic eigenvalues of $V$. To simplify
the notation we now introduce:\begin{subequations}
\begin{equation}
z_{l}=\begin{cases}
i\gamma_{l} & l=1,\ldots,N\\
-i\gamma_{l-N} & l=N+1,\ldots,2N
\end{cases},
\end{equation}
\begin{equation}
\left|z_{l}\right\rangle =\begin{cases}
\left|\gamma_{l}^{+}\right\rangle  & l=1,\ldots,N\\
\left|\gamma_{l-N}^{-}\right\rangle  & l=N+1,\ldots,2N
\end{cases}.
\end{equation}
\end{subequations}Obviously $\mathcal{V}\left|z_{l}\right\rangle =z_{l}\left|z_{l}\right\rangle $
for $l=1,\ldots,2N$. 

In terms of the covariance
matrix, stabilizability by definition means $\dot{V}=0$, or more precisely
\begin{equation}
AJ^{T}\tilde{V}+\tilde{V}^{T}JA^{T}+J\left(\textrm{Re}C^{\dagger}C\right)J^{T}=0.\label{stabilizab}
\end{equation}
This simply follows from the evolution equation with the covariance
matrix appropriately replaced by $\tilde{V}$. In Appendix \ref{AppA}
we prove that if the covariance matrix is invertible and the spectrum
of $\mathcal{V}$ is non-degenerate, then the choice\begin{subequations}\label{Hamil}
\begin{equation}
G=J^{T}V^{1/2}\left(\sum_{l'\neq l}\frac{\mathcal{D}_{l'l}}{z_{l'}-z_{l}}\left|z_{l'}\right\rangle \left\langle z_{l}\right|\right)V^{1/2}J,
\end{equation}
\begin{equation}
\mathcal{D}_{l'l}=\left\langle \tilde{z}_{l'}\right|\left(z_{l}+z_{l'}\right)\textrm{Im}C^{\dagger}C+\textrm{Re}C^{\dagger}C\left|\tilde{z}_{l}\right\rangle ,
\end{equation}
with $\left|\tilde{z}_{l}\right\rangle =J^{T}V^{-1/2}\left|z_{l}\right\rangle $
\end{subequations} solves (\ref{stabilizab}), given that the conditions
(\ref{CriteriaMain}) are taken into account. In other words, Eqs.
\ref{Hamil} specify the proper quadratic Hamiltonian present in the definition
of stabilizability. 

Let us remark that, for the sake of transparency, in (\ref{HamiltonianG}) we omit terms linear in $\hat{\xi}$. The most general quadratic Hamiltonian can be written (up to irrelevant constant factors) as $\hat{H}=\frac{1}{2}(\hat{\xi}-\xi_0)^{T}G(\hat{\xi}-\xi_0)$, which basically accounts for a shift/redefinition of the vacuum in phase space. The above discussed stabilization of coherent states under damping (\ref{Eq:damping_dissipator}) can then be understood as the stabilization of shifted ground states, taking into account that the dissipator (\ref{Eq:damping_dissipator}) is, when evaluated for coherent states $\ket{\alpha}$, invariant under the transformation $\hat{a} \rightarrow \hat{a}-\alpha_0$, $\mathcal{D}_{a-\alpha_0}(\ketbra{\alpha}{\alpha}) = \mathcal{D}_{a}(\ketbra{\alpha}{\alpha})$. Note that this perspective implies a diagonal, but necessarily non-vanishing matrix $G$ in (\ref{HamiltonianG}), which then also introduces terms quadratic in the coordinates (in contrast to the stabilizing Hamiltonian derived above, which is linear in the coordinates). This indicates that stabilizing Hamiltonians are, in general, not unique, as one is free to modify the Hamiltonian which stems from Eq. \ref{Hamil} (for the case of the coherent state it gives $G=0$) by adding terms which feature the stabilizable states as eigenstates--in the considered case, e.g., $\hat{H}= \hat{a}^{\dagger} \hat{a}$.

\section{Examples} 

\subsection{Single damped mode}

As a simple but at the same time quite instructive example, we further elaborate on the Gaussian states of a single damped mode ($N=1$), characterized by one ($M=1$) Lindblad operator
\begin{align} \label{Eq:single_mode_damping_Lindbladian}
\hat{L} = \sqrt{\gamma} \, \hat{a} .
\end{align}
As already shown in Sec. \ref{Sec1B}, this dissipator allows all (pure) coherent states to be stabilized.
In the following, we are interested in the general case, which covers, e.g., squeezed and all kinds of mixed states.

Rewriting the Lindblad operator (\ref{Eq:single_mode_damping_Lindbladian}) in terms of its quadratures, say, the position $\hat{x}$ and the momentum $\hat{p}$, $\hat{a} = (\hat{x}/x_0 + i \, \hat{p} \, x_0)/\sqrt{2}$, we obtain the dissipation matrix
\begin{align}
C = \sqrt{\frac{\gamma}{2}} \, (x_0^{-1} , i \, x_0) .
\end{align}
Note that this step introduced a characteristic length scale $x_0$. In the case of the standard harmonic oscillator, it is determined by the parameters of the Hamiltonian, $x_0 = (\omega m)^{-1/2}$; here, however, we leave the Hamiltonian \textit{a priori} undetermined. For the evaluation of the stabilizability conditions, we determine
\begin{align}\label{CdC}
C^{\dagger} C = \frac{\gamma}{2} \left[ \left( \begin{array}{cc} x_0^{-2} & 0 \\ 0 & x_0^2 \end{array} \right) + i \left( \begin{array}{cc} 0 & 1 \\ -1 & 0 \end{array} \right) \right] .
\end{align}

Since $N=1$, the Gaussian stabilizability conditions (\ref{CriteriaMain}) provide us with two constraints on the possible stabilizable states. As $\gamma$ is a multiplicative factor in (\ref{CdC}), these constraints are $\gamma$-independent.  Evaluated for $n=1$, the corresponding condition reads
\begin{align}
V_{x p} - V_{p x} = 0 ,
\end{align}
which is automatically satisfied, as the covariance matrix is symmetric by definition. The constraint for $n=1$ assumes this simple form due to the fact that the imaginary part of $C^{\dagger} C$ is proportional to $J$.

\begin{figure}[htb]
	(a) \phantom{aaaaaaaaaaaaaaaaaaaaaa} (b) \phantom{aaaaaaaaaaaaaaaaaaaaaa} \\ \vspace{2mm}
	\includegraphics[width=0.45\columnwidth]{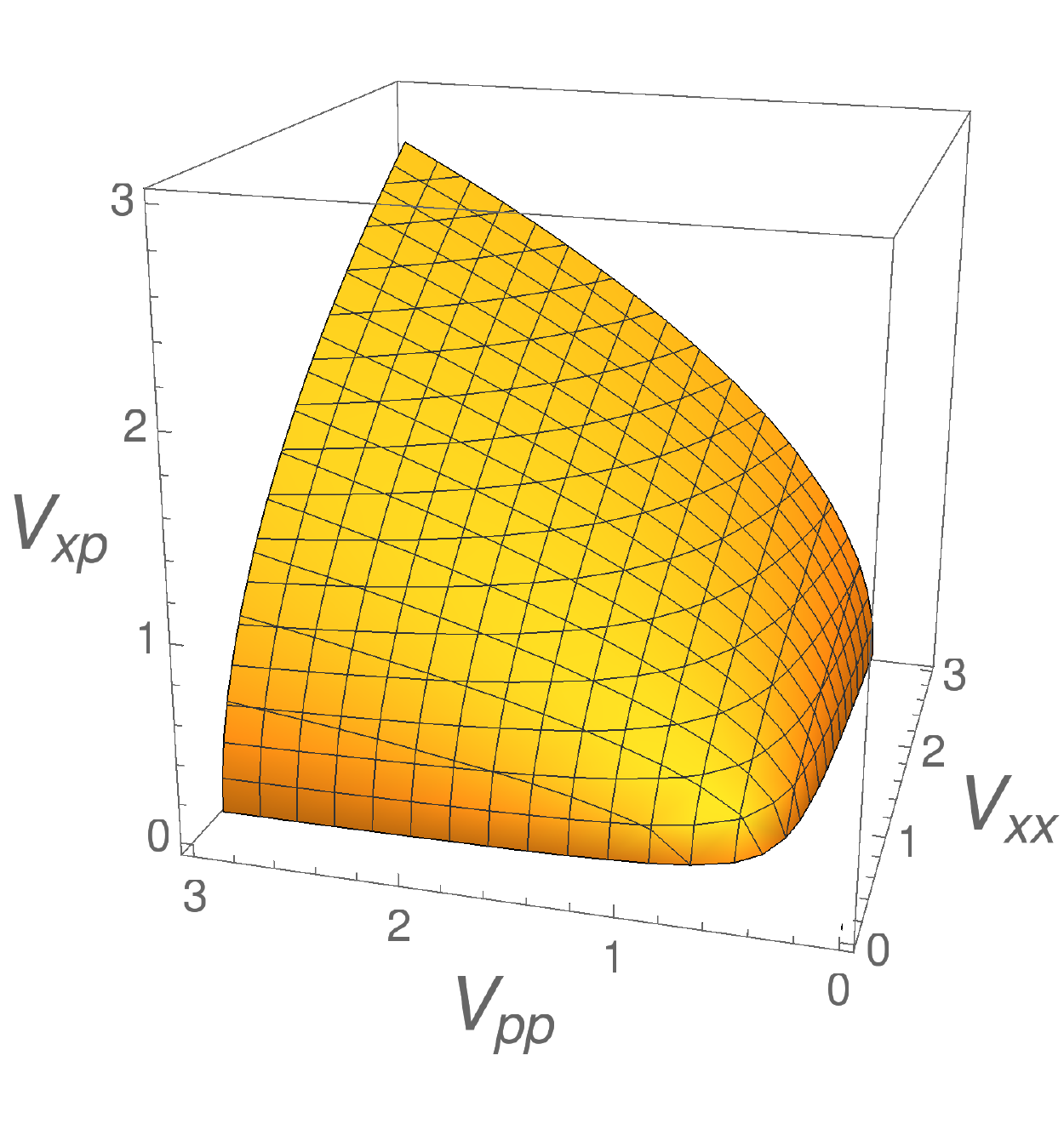}
	\phantom{aa}
	\includegraphics[width=0.45\columnwidth]{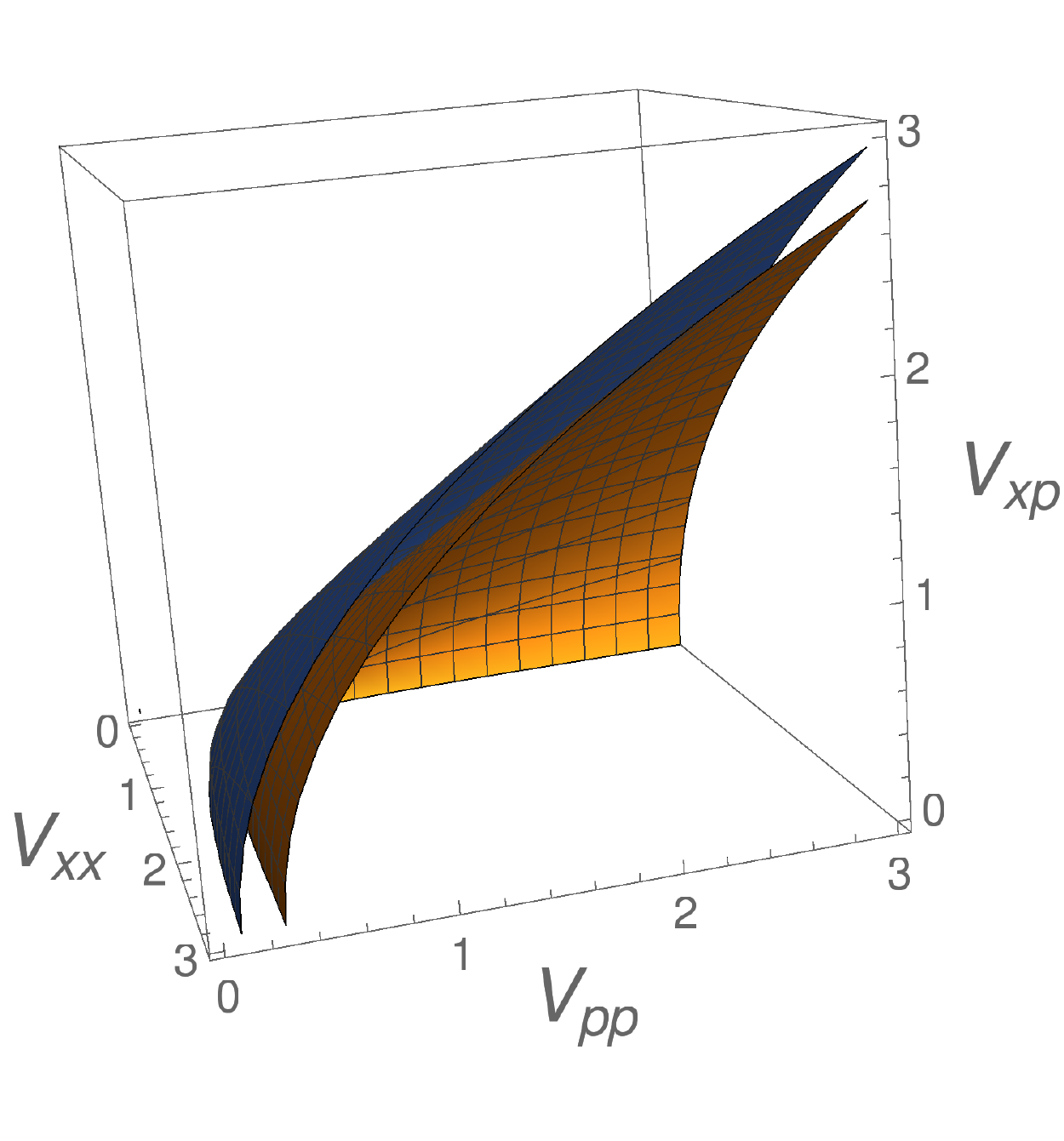}
	(c) \phantom{aaaaaaaaaaaaaaaaaaaaaaaaaaaaaaaaaaaaaaaaaaaaaaa} \\
	\vspace{2mm}
	\includegraphics[width=0.8\columnwidth]{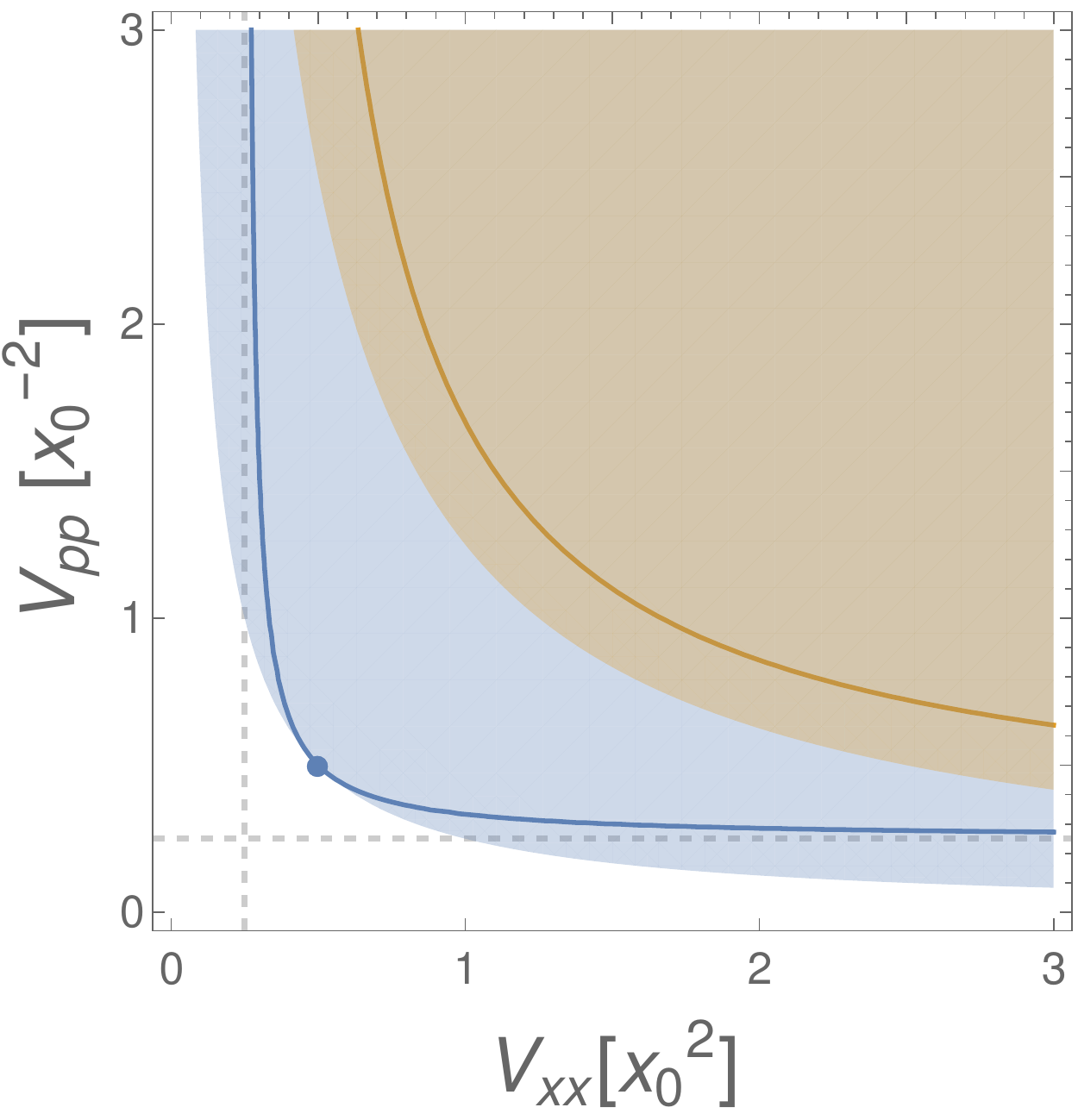}
	\caption{\label{Fig:squeezed_state_stabilization} (Color online) Set of Gaussian stabilizable states for a single damped mode. (a) The $n=2$ condition (\ref{Eq:single_mode_damping_stabilizability_condition}) describes a segment of a two-sheeted hyperboloid [units as in (c)]. (b) All stabilizable states (orange sheet) are consistent with the generalized uncertainty relation (\ref{Eq:single_mode_generalized_uncertainty_relation}). Shown is the set of states satisfying (\ref{Eq:single_mode_generalized_uncertainty_relation}) with equality (blue sheet). (c) Stabilizable states for fixed $V_{x p} = 0$ (blue line) and $V_{x p} = 1$ (orange line), respectively. Squeezing is limited to $V_{x x} = x_0^2/4$ and $V_{p p} = x_0^{-2}/4$, respectively (gray dashed lines). All stabilizable states respect the generalized uncertainty relation (\ref{Eq:single_mode_generalized_uncertainty_relation}) (blue and orange area, respectively). The only pure stabilizable states are the coherent states with $V_{x x} = x_0^2/2$, $V_{p p} = x_0^{-2}/2$, and $V_{x p} = 0$ (blue dot).}
\end{figure}

This leaves us with the $n=2$ constraint, which evaluates as $\Omega(x,p)=$
\begin{align} \label{Eq:single_mode_damping_stabilizability_condition}
 x_0^{-2} V_{x x} + x_0^2 V_{p p} - 4 V_{x x} V_{p p} +2 (V_{x p}^2 + V_{p x}^2)= 0 .
\end{align}
This stabilizability condition, which is also depicted in Fig.~\ref{Fig:squeezed_state_stabilization}, describes the segment of a two-sheeted hyperboloid in the $(+,+,+)$-octant of the parameter space of the covariance matrix. This can easily be seen by rewriting Eq.~(\ref{Eq:single_mode_damping_stabilizability_condition}) in terms of the variables $x =x_0^{-2}  V_{x x} + x_0^{2}  V_{p p} - 1/2$, $y =x_0^{-2}  V_{x x} -x_0^{2} V_{p p}$, and $z = 2 V_{x p}$, which then takes the form $x^2 - y^2 - z^2 = 1/4$. Importantly, all stabilizable states are consistent with the generalized uncertainty relation $V+i J/2 \geq 0$ \cite{Yamamoto1, Pirandola2009correlation}, here
\begin{align} \label{Eq:single_mode_generalized_uncertainty_relation}
V_{x x} V_{p p} - V_{x p}^2 \geq  \frac{1}{4} ,
\end{align}
as shown in Fig.~\ref{Fig:squeezed_state_stabilization}(b).

The stabilizability condition (\ref{Eq:single_mode_damping_stabilizability_condition}) allows us to identify possible stabilizable Gaussian states. Squeezed states being an important quantum resource, e.g. in quantum metrology \cite{Caves1981quantum}, we now quantify the possibility to stabilize squeezed Gaussian states, with either $V_{x x}$ or $V_{p p}$ suppressed.

For the sake of clarity, we consider in the following the squeezing of the position variance. To determine the minimum possible position uncertainty, we rewrite Eq.~(\ref{Eq:single_mode_damping_stabilizability_condition}) as
\begin{align}
V_{x x} = \frac{x_0^2 V_{p p} + 4 V_{x p}^2}{4 V_{p p} - x_0^2} ,
\end{align}
excluding $V_{p p} = x_0^{-2}/4$. It is now easy to see that the minimum variance is determined by $V_{x p} = 0$ and $V_{p p} \rightarrow \infty$, which yields 
\begin{align}
V_{x x} = \frac{x_0^2}{4} ,
\end{align}
i.e., we find that the position variance is limited to finite squeezing. In the limit $V_{x x} \rightarrow x_0^2/4$, the corresponding momentum variance diverges. We remark that a similar result is obtained for the optimal squeezing of the momentum variable.

The optimal stabilizable state in terms of minimal position variance, characterized by $V_{x x} \rightarrow x_0^2/4$, $V_{p p} \rightarrow \infty$, and $V_{x p} = 0$, is highly mixed, as reflected by the purity $p = \tr[\rho^2] = 1/\sqrt{2^{2 N} {\rm det}(V)}$ \cite{Yamamoto1}, which vanishes in the limit:
\begin{align}
p = \frac{1}{2 \sqrt{V_{x x} V_{p p} - V_{x p}^2}} \rightarrow 0 .
\end{align}
If we restrict us to the stabilization of pure states, with $p=1$, we find that the only pure stabilizable state is characterized by $V_{x x} = x_0^2/2$, $V_{p p} = x_0^{-2}/2$, and $V_{x p} = 0$, as confirmed by inspection of Fig.~\ref{Fig:squeezed_state_stabilization}. This is the ground state of a harmonic oscillator with characteristic length $x_0$. More generally, if we allow for shifted ground states in (\ref{HamiltonianG}) and recall the shift invariance of the dissipator (\ref{Eq:damping_dissipator}), this recovers the above discussed stabilizability of coherent states. Indeed, all coherent states share the covariance matrix of the ground state, $V = {\rm diag}(x_0^2/2, x_0^{-2}/2)$, which, as one can easily verify, satisfies the stabilizability condition (\ref{Eq:single_mode_damping_stabilizability_condition}).

\subsection{Two locally damped modes}

For a second example, we consider two modes with local damping. This describes a generic situation for several quantum technologies employing entangled Gaussian states, e.g., CV quantum teleportation with Gaussian resource states \cite{Braunstein1998teleportation}. Local damping acts detrimentally on entangled states, and therefore we investigate to what extent entangled Gaussian states can be uphold with quadratic Hamiltonians.

We now have two Lindblad operators
\begin{align}
\hat{L}_i = \sqrt{\gamma_i} \hat{a}_i ,
\end{align}
resulting in the dissipation matrix
\begin{align}
C = \frac{1}{\sqrt{2}}\left(\begin{array}{cccc}
\sqrt{\gamma_{1}}x_{0}^{-1} & 0 & i\sqrt{\gamma_{1}}x_{0} & 0\\
0 & \sqrt{\gamma_{2}}x_{0}^{-1} & 0 & i\sqrt{\gamma_{2}}x_{0}
\end{array}\right) ,
\end{align}
which, in turn, gives rise to
\begin{align}
C^{\dagger} C =  \frac{1}{2} \tilde\gamma \left[ \left( \begin{array}{cc} x_0^{-2} \mathbb{1}_2 & 0 \\ 0 & x_0^2 \, \mathbb{1}_2 \end{array} \right) + i J \right] ,
\end{align}
with $\tilde \gamma$ being a four-dimensional diagonal matrix $( \gamma_1, \gamma_2,\gamma_1, \gamma_2)$. Note that, in principle, one is free to chose different length scales ($x_0$) for each mode, which we avoid here for the sake of simplicity and without sacrificing significance. Evaluating the stabilizability conditions (\ref{CriteriaMain}) then yields, for $n=1$,
\begin{align}
\gamma_{1}(V_{x_1 p_1} - V_{p_1 x_1}) + \gamma_{2} (V_{x_2 p_2} - V_{p_2 x_2}) = 0 ,
\end{align}
which, again, is automatically satisfied due to the symmetry of the covariance matrix. The $n=2$ condition, on the other hand, reads
\begin{align} \label{Eq:two_damped_modes_stabilizability_condition}
\gamma_{1}\Omega(x_1,p_1)+\gamma_{2}\Omega(x_2,p_2) - 4(\gamma_{1}+\gamma_{2}) {\rm det} \, V_{12}= 0 ,
\end{align}
with  the first two terms reflecting the local $n=2$ stabilizability conditions of the two single modes ($\Omega$ has been defined above Eq.~\ref{Eq:single_mode_damping_stabilizability_condition}), and the matrix
 \begin{equation}
V_{12} = \left( \begin{array}{cc} V_{x_1 x_2} & V_{x_1 p_2} \\ V_{p_1 x_2} & V_{p_1 p_2} \end{array} \right),
\end{equation}
encoding non-local correlations. As was shown in \cite{Simon2000peres}, Gaussian states can only be entangled if ${\rm det} \, V_{12} < 0$.

We thus obtain the interesting result that, in the presence of local damping, entanglement of stabilizable Gaussian states is completely determined by their local properties, weighted by the damping strengths, as per the relation:

\begin{equation}
{\rm det} \, V_{12} = \frac{\gamma_{1}}{4(\gamma_{1}+\gamma_{2})}\Omega(x_1,p_1)+\frac{\gamma_{2}}{4(\gamma_{1}+\gamma_{2})}\Omega(x_2,p_2).
\end{equation}

Note that, in principle, there are two more stabilizability conditions, corresponding to $n=3$ and $n=4$. However, in our discussion we restrict to  $n=2$ as a necessary requirement, and leave a detailed discussion of the complete geometry of two-mode stabilizable Gaussian states for the future.

The necessary condition (\ref{Eq:two_damped_modes_stabilizability_condition}) alone can already be used to gain insight into the restrictions on the stabilizable states. For example, evaluating (\ref{Eq:two_damped_modes_stabilizability_condition}) with an ansatz for pure Einstein-Podolsky-Rosen (EPR) correlated states,
\begin{align}
\Psi(p_{\rm cm}, x_{\rm rel}) = \frac{1}{\sqrt{2 \pi \sigma_{p, {\rm cm}} \sigma_{x, {\rm rel}}}} e^{-\frac{p_{\rm cm}^2}{4 \sigma_{p, {\rm cm}}^2}} e^{-\frac{x_{\rm rel}^2}{4 \sigma_{x, {\rm rel}}^2}} ,
\end{align}
one obtains (for simplicity, we assume in the following $\gamma_{1} = \gamma_{2}$)
\begin{align}
\frac{1}{2} x_0^{-2} \sigma_{p, {\rm cm}}^{-2} + \frac{1}{2} x_0^{-2} \sigma_{x, {\rm rel}}^{2} + \frac{1}{2} x_0^{2} \sigma_{p, {\rm cm}}^{2} + \frac{1}{2} x_0^{2} \sigma_{x, {\rm rel}}^{-2} = 2 ,
\end{align}
which is only satisfied if $\sigma_{p, {\rm cm}} = x_0^{-1}$ and $\sigma_{x, {\rm rel}} = x_0$. This, however, implies that ${\rm det} \, V_{12} = 0$. In other words, the only pure EPR state which can be stabilized is separable. Indeed, this describes the ground/vacuum state of the two-mode system, in agreement with a dark-state analysis of the dissipator.

\section{Discussion} 

We investigated stabilizability for Gaussian CV systems. To this end, we derived, based on the invariance of the symplectic eigenvalues of the covariance matrix, the necessary and sufficient stabilizability criteria (\ref{CriteriaMain}). These criteria impose extra constraints on the covariance matrix, offering a geometric intuition and a playground for optimization. Working with the set of stabilizable states characterized that way, one can, for instance, look for minimal/maximal values assumed by quantities of potential relevance, such as logarithmic negativity \cite{LogNeg}.

For non-Gaussian states, higher-order moments come also into play. However, the covariance matrix follows the same evolution as for Gaussian states and can thus be controlled in the same way.
This may, e.g., be useful to discuss entanglement preservation and/or detection in the non-Gaussian case. For example, covariance-based entanglement criteria (e.g. \cite{Simon2000peres}), which are necessary and sufficient for Gaussian states, remain useful in non-Gaussian scenarios, where they still provide necessary conditions. One can thus consider a restricted problem, in which stabilizability is only imposed on the covariance matrix, letting higher (non-trivial) moments evolve.

In principle, one could generalize the results for the covariance matrix to the covariant moments of arbitrary order \cite{Mukunda}, since the moment-matrices are subject to the same kind of $Sp(2N,\mathbb{R})$ symmetry (but according to different representations of this group).
However, since stabilization of the covariance matrix already absorbs the adjustable parameters in Gaussian Hamiltonians, one should in general not expect that such stabilization can be achieved by quadratic means. This expectation is also corroborated by a phase-space argument: Since the Wigner function of an initial state $W_{0}\left(\xi\right)$ evolves under (\ref{evol}) such that $W\left(\xi,t\right)$ is the convolution of $W\left(\xi,0\right)\equiv W_{0}\left(\xi\right)$ with a given (time-dependent) Gaussian \cite{Wigner1}, it seems impossible to satisfy the requirement $W\left(\xi,t\right)=W_{0}\left(\xi\right)$ unless $W_{0}\left(\xi\right)$ is also Gaussian.

In the case of non-Gaussian states one thus needs to consider broader families of Hamiltonians and to develop an extended theoretical description. An immediate candidate is a quadratic Hamiltonian with the addition of a non-linear Kerr interaction, which is, e.g., successful in (approximately) stabilizing cat states \cite{Mamaev}, though the role played by the environment is different in \cite{Mamaev}.
Finally, let us note in passing that an interesting open problem for the future is to relate the approach based on stabilizability with quantum error-correcting codes utilizing CV systems \cite{ECCodes}, also beyond the regime of the Gaussian states.

\acknowledgments

Financial support from the National Science Center, Poland, under
Grant No. 2014/13/D/ST2/01886 is gratefully acknowledged. 

\appendix\section{Derivation of the Hamiltonian (\ref{Hamil})}\label{AppA}

Since $\tilde{V}^{T}=-VJ$ and $\mathcal{V}^{\dagger}=\mathcal{V}^{T}=-\mathcal{V}$
we can easily see that
\begin{equation}
\mathcal{V}=V^{1/2}\tilde{V}V^{-1/2},\qquad\mathcal{V}^{\dagger}=V^{-1/2}\tilde{V}^{T}V^{1/2}.
\end{equation}
Thus\begin{subequations}\label{EigEq}
\begin{equation}
\tilde{V}V^{-1/2}\left|z_{l}\right\rangle =z_{l}V^{-1/2}\left|z_{l}\right\rangle ,\label{eigenEq}
\end{equation}
and since $\left\langle z_{l}\right|\mathcal{V}^{\dagger}=z_{l}^{*}\left\langle z_{l}\right|=-z_{l}\left\langle z_{l}\right|$
we have
\begin{equation}
\left\langle z_{l}\right|V^{-1/2}\tilde{V}^{T}=-z_{l}\left\langle z_{l}\right|V^{-1/2}.\label{eigenEq-1}
\end{equation}
\end{subequations}One needs to emphasize, as was already assumed,
that the covariance matrix needs to be invertible, so that $V^{-1/2}$
is well defined.

If we now multiply Eq. \ref{stabilizab} from the left and from the
right by $V^{-1/2}$, and further, as shown in (\ref{stabDef}), test
the resulting expression with the help of the orthonormal basis $\left\{ \left|z_{l}\right\rangle \right\} $
and their properties (\ref{EigEq}), we obtain
\begin{equation}
\left(z_{l}-z_{l'}\right)\left\langle \tilde{z}_{l'}\right|G\left|\tilde{z}_{l}\right\rangle +\mathcal{D}_{l'l}=0.\label{major}
\end{equation}
Clearly, if $l\neq l'$, then also $z_{l}-z_{l'}\neq0$ and we can
solve the above equation with respect to $G$. One can check by a
direct substitution that the solution is given by (\ref{Hamil}).

We thus only need to check what happens with the condition (\ref{major})
when $l=l'$. In this special case it is required that $\mathcal{D}_{ll}=0$
for all $l=1,\ldots,2N$. It is expected \cite{Sauer}, that these
conditions shall be satisfied by virtue of the criteria (\ref{CriteriaMain}).
To prove this is true, we first observe that $\tilde{V}^{-1}=V^{-1}J^{T}$,
and then utilize the relation

\begin{equation}
V^{-1}=V^{-1/2}\sum_{l}\left|z_{l}\right\rangle \left\langle z_{l}\right|V^{-1/2},\label{Vm1}
\end{equation}
valid because 
\begin{equation}
\sum_{l}\left|z_{l}\right\rangle \left\langle z_{l}\right|=\1_{2N},
\end{equation}
to show
\begin{equation}
\textrm{Tr}\left[\left(\textrm{Im}C^{\dagger}C\right)J\tilde{V}^{n}\right]=\sum_{l}z_{l}^{n+1}\left\langle \tilde{z}_{l}\right|\left(\textrm{Im}C^{\dagger}C\right)\left|\tilde{z}_{l}\right\rangle .
\end{equation}
To this end, one simply needs to replace $\tilde{V}^{n}$ by $\tilde{V}^{n+1}V^{-1} J^T$,
and use (\ref{Vm1}) together with the relation (\ref{eigenEq}) applied
$n+1$ times, and basic properties of the $J$ matrix. In almost exactly
the same way ($n$ replaced by $n-1$) one obtains 
\begin{equation}
\textrm{Tr}\left[\left(\textrm{Re}C^{\dagger}C\right)J\tilde{V}^{n-1}\right]=\sum_{l}z_{l}^{n}\left\langle \tilde{z}_{l}\right|\left(\textrm{Re}C^{\dagger}C\right)\left|\tilde{z}_{l}\right\rangle .
\end{equation}
Adding the last equation to the former one multiplied by $2$, one
arrives at 
\begin{equation}
0=\sum_{l}z_{l}^{n}\mathcal{D}_{ll},\label{Vander}
\end{equation}
for all $n=1,\ldots,2N$, by recognizing that the left hand side of
such a sum must vanish due to the criteria (\ref{CriteriaMain}).
Finally, the Vandermonde-matrix argument \cite{Sauer} applied to
Eq. \ref{Vander} proves the desired equality $\mathcal{D}_{ll}=0$.
To be more specific, the Vandermonde matrix $\mathfrak{M}_{ln}=z_{l}^{n-1}$
is invertible, so that the vector $z_{l}\mathcal{D}_{ll}$ must be
equal to $0$ (the eigenvalues $z_{l}$ are non-zero by assumption).


\begin{thebibliography}{References}
\bibitem{FTQC} D. Gottesman, \emph{Designing Quantum Memories with
Embedded Control: Photonic Circuits for Autonomous Quantum Error Correction},
arXiv:0904.2557v1. 

\bibitem{Grav} B.\LyXThinSpace P. Abbott et al., \emph{Observation
of Gravitational Waves from a Binary Black Hole Merger}, Phys. Rev.
Lett. \textbf{116}, 061102 (2016).

\bibitem{Mandel1995optical} L. Mandel and E. Wolf, \emph{Optical coherence and quantum optics},
Cambridge University Press, 1995.

\bibitem{Walls2007quantum} D. Walls and G. J. Milburn, \emph{Quantum optics},
Springer Science and Business Media, 2007.

\bibitem{Wang2007quantum} X.-B. Wang, T. Hiroshima, A. Tomita and M. Hayashi, \emph{Quantum
information with Gaussian states}, Physics Reports \textbf{448}, 1 (2007).

\bibitem{Weedbrook2012gaussian} C. Weedbrook, S. Pirandola, R. Garc{\'\i}a-Patr{\'o}n, N. J. Cerf, T. C. Ralph, J. H. Shapiro and S. Lloyd, \emph{Gaussian quantum information}, Rev. Mod. Phys. \textbf{84}, 621 (2012).

\bibitem{Braunstein1998teleportation} S. L. Braunstein and H. J. Kimble, \emph{Teleportation
of Continuous Quantum Variables}, Phys. Rev. Lett. \textbf{80}, 869 (1998).

\bibitem{Liuzzo2017optimal} P. Liuzzo-Scorpo, A. Mari, V. Giovannetti and G. Adesso, \emph{Optimal Continuous Variable Quantum Teleportation with Limited Resources}, Phys. Rev. Lett. \textbf{119}, 120503 (2017).

\bibitem{Furusawa1998unconditional} A. Furusawa, J. L. S{\o}rensen, S. L. Braunstein, C. A. Fuchs, H. J. Kimble and E. S. Polzik, \emph{Unconditional quantum teleportation}, Science \textbf{282}, 706 (1998).

\bibitem{Kheruntsyan2005einstein} K. V. Kheruntsyan, M. K. Olsen and P. D. Drummond, \emph{Einstein-Podolsky-Rosen Correlations via Dissociation of a Molecular Bose-Einstein Condensate}, Phys. Rev. Lett. \textbf{95}, 150405 (2005).

\bibitem{Gneiting2013nonlocal} C. Gneiting and K. Hornberger, \emph{Nonlocal Young tests with Einstein-Podolsky-Rosen-correlated particle pairs}, Phys. Rev. A \textbf{88}, 013610 (2013).

\bibitem{Sauer} S. Sauer, C. Gneiting, and A. Buchleitner, \emph{Optimal
Coherent Control to Counteract Dissipation}, Phys. Rev. Lett. \textbf{111},
030405 (2013).

\bibitem{Sauer2014stabilizing} S. Sauer, C. Gneiting and A. Buchleitner, \emph{Stabilizing
entanglement in the presence of local decay processes}, Phys. Rev. A \textbf{89}, 022327 (2014).

\bibitem{Addis2016problem} C. Addis, E.-M. Laine, C. Gneiting and S. Maniscalco, \emph{Problem
of coherent control in non-Markovian open quantum systems}, Phys. Rev. A \textbf{94}, 052117 (2016).

\bibitem{Mancini} S. Mancini, \emph{Markovian feedback to control
continuous-variable entanglement}, Phys. Rev. A \textbf{73}, 010304R
(2006).

\bibitem{Mancini2} S. Mancini and H. M. Wiseman, \emph{Optimal control
of entanglement via quantum feedback}, Phys. Rev. A \textbf{75}, 012330
(2007).

\bibitem{Yamamoto1} K. Koga and N. Yamamoto, \emph{Dissipation-induced
pure Gaussian state}, Phys. Rev. A \textbf{85}, 022103 (2012).

\bibitem{Yamamoto2} N. Yamamoto, \emph{Pure Gaussian state generation
via dissipation: a quantum stochastic differential equation approach},
Phil. Trans. R. Soc. A \textbf{370}, 5324 (2012). 

\bibitem{Serafini} A. Serafini, \textit{Quantum Continuous Variables: A Primer of Theoretical Methods}, (CRC Press), 2017.

\bibitem{SPDC} S. P. Walborn, C. H. Monken, S. P{\'a}dua, and P. H. S.
Ribeiro, \emph{Spatial correlations in parametric down-conversion},
Phys. Rep. \textbf{495}, 87 (2010).

\bibitem{Pirandola2009correlation} S. Pirandola, A. Serafini and S. Lloyd, \emph{Correlation matrices
of two-mode bosonic systems}, Phys. Rev. A \textbf{79}, 052327 (2009).


\bibitem{Caves1981quantum} C. M. Caves, \emph{Quantum-mechanical noise in an
interferometer}, Phys. Rev. D \textbf{23}, 1693 (1981).

\bibitem{Simon2000peres} R. Simon, \emph{Peres-Horodecki Separability Criterion
for Continuous Variable Systems}, Phys. Rev. Lett. \textbf{84}, 2726 (2000).

\bibitem{LogNeg} G. Adesso, A. Serafini, and F. Illuminati, {\emph Extremal entanglement
and mixedness in continuous variable systems},
Phys. Rev. A {\bf 70}, 022318 (2004).

\bibitem{Mukunda} J. Solomon Ivan, N. Mukunda, and R. Simon, \emph{Moments
of non-GaussianWigner distributions and a generalized uncertainty
principle: I. The single-mode case}, J. Phys. A: Math. Theor. \textbf{45}
195305 (2012).

\bibitem{Wigner1} O. Brodier and A. M. Ozorio de Almeida, \emph{Symplectic
evolution of Wigner functions in Markovian open systems}, Phys. Rev.
E \textbf{69}, 016204 (2004).

\bibitem{Mamaev} M. Mamaev, L. C. G. Govia, and A. A. Clerk, \emph{Dissipative stabilization of entangled cat states using a driven Bose-Hubbard dimer}, 
arXiv:1711.06662

\bibitem{ECCodes} D. Gottesman, A. Kitaev, and J. Preskill, \emph{Encoding a qubit in an oscillator}, Phys. Rev. A {\bf 64}, 012310 (2001).



\end{thebibliography}
\end{document}